\documentclass[twocolumn,prb,tightenlines,superscriptaddress,showpacs]{revtex4-2}

\usepackage{amsmath}
\usepackage{amssymb,amsfonts,latexsym}
\usepackage{bm}
\usepackage[mathcal]{euscript}
\usepackage{graphicx}
\usepackage{epsfig}
\usepackage{color}

\newcommand{\be}{\begin{equation}}
\newcommand{\ee}{\end{equation}}
\newcommand{\ve}{\varepsilon}
\newcommand{\la}{\langle}
\newcommand{\ra}{\rangle}

\newcommand{\modif}[1]{\textcolor{black}{#1}}
\newcommand{\modifnew}[1]{\textcolor{black}{#1}}

\begin{document}

\title{Tailoring the overlap distribution in driven mean-field spin models}

\author{Laura Guislain}
\affiliation{Univ.~Grenoble Alpes, CNRS, LIPhy, 38000 Grenoble, France}

\author{Eric Bertin}
\affiliation{Univ.~Grenoble Alpes, CNRS, LIPhy, 38000 Grenoble, France}

\date{\today}

\begin{abstract}
In a statistical physics context, inverse problems consist in determining microscopic interactions such that a system reaches a predefined collective state.
A complex collective state may be prescribed by specifying the overlap distribution between microscopic configurations, a notion originally introduced in the context of disordered systems like spin-glasses.
We show that in spite of the absence of disorder, nonequilibrium spin models exhibiting spontaneous magnetization oscillations provide a benchmark to prescribe a non-trivial overlap distribution with continuous support, qualitatively analogous to the ones found in disordered systems with full replica symmetry breaking. The overlap distribution can be explicitly tailored to take a broad range of predefined shapes by monitoring the spin dynamics.
The presence of a non-trivial overlap distribution is traced back to an average over infinitely many pure states, a feature shared with spin-glasses, \modif{although the structure of pure states is here much simpler}.
\end{abstract}

\maketitle

\section{Introduction}
Finding microscopic interactions or dynamical rules such that a system composed of many interacting subunits reaches a predefined macroscopic state is often challenging \cite{cranmer2020}, as it may involve exploring a potentially large parameter space. 
Such computationally demanding problems are called inverse problems, and their interest has been renewed by the recent development of powerful machine learning techniques \cite{cranmer2020,nguyen2017}.
Inverse problems are very relevant in particular when searching for a microscopic model able to describe real-world data \cite{nguyen2017}.
For example, interaction networks have been reconstructed from measured data in the case of starling flocks \cite{mora2016,bialek2012}, neural networks \cite{schneidman2006,cocco2009} or gene regulatory networks \cite{lezon2006,locasale2009,molinelli2013}.
Inverse problems are also currently explored in the context of active matter, where an emergent research trend consists in using decentralized learning procedures to optimize individual parameters to reach a target collective behavior \cite{durve2020,falk2021,vansaders2023,nasiri2023,devereux2023,benzion2023}.

The collective state of an assembly of interacting entities may be characterized by global observables, like polar or nematic order parameters in an active matter context for instance. Yet, a finer characterization of the collective structure may be obtained from the overlap distribution of microscopic configurations. The overlap $q$ of two configurations is a measure of their similarity, typically $q \approx 1$ for similar configurations, while $q\approx 0$ when two configurations are picked up at random.
The probability distribution of overlaps has been introduced as a useful characterization of the spin-glass phase, to deal with the absence of a visible order \cite{mezard_spin_1987,carpentier2008,wittmann2014}. It has later proved to be an efficient characterization of the glassy state in supercooled liquids \cite{guiselin2022,berthier2016,berthier2013}, random lasers and disordered optical media \cite{leuzzi2023,ghofraniha2015}, constraint satisfaction problems \cite{krabbe2023,mezard2005}, neural networks \cite{montemurro2000,gyorgyi2000}, population dynamics \cite{altieri2021,manzo1994} and directed polymer models \cite{hartmann2022}.
%
Quite importantly, the overlap distribution is non-trivial only for `complex' phases of matter, like glassy phases that exhibit replica symmetry breaking \cite{mezard_spin_1987}. For ordinary phases of matter like liquid or gas phases, the overlap distribution $P(q)$ reduces to a delta peak at $q=0$ in the thermodynamic limit, as if configurations were picked completely at random. For ordered phases, $P(q)$ is a sum of delta peaks related by the broken symmetry (e.g., two symmetric delta peaks for a ferromagnetic state).
\modif{In the specific case of the critical XY model, an overlap distribution that does not reduce to a sum of delta peaks has been reported, at least for finite sizes \cite{Berthier2001}, and can be traced back to the $O(2)$-symmetry of the model.}
However, imposing a non-trivial prescribed overlap distribution is in general a challenging task, as one needs both to generate enough complexity in the state of the system, and to tame it so as to impose the prescribed overlap distribution, which typically does not correspond to the one spontaneously observed in the system.

In this paper, we argue that in spite of the absence of disorder, driven mean-field spin models exhibiting spontaneous magnetization oscillations are characterized by a non-trivial overlap distribution with a continuous support. We show that this overlap distribution can be explicitly tailored to take a broad range of predefined shapes by monitoring the spin dynamics, for instance through a magnetization-dependence of the spin-flip frequency.
In addition, the physical origin of the overlap distribution is traced back to an average over infinitely many pure states, \modif{leading to a partial and} qualitative analogy with mean-field spin-glass models.

\section{Generic mean-field spin model}

We consider a generic mean-field spin model composed of $N$ spins $s_i=\pm 1$, endowed with a stochastic spin-flip dynamics that breaks detailed balance.
The model also typically includes auxiliary variables to allow for spontaneous temporal oscillations of the magnetization $m=\frac{1}{N} \sum_{i=1}^N s_i$
in a far-from-equilibrium regime \cite{guislain_nonequil2023,guislain_discontinuous2023}
--see also \cite{collet_macroscopic_2014,collet_rhythmic_2016,collet_effects_2019,de_martino_oscillations_2019,martino_feedback2019,daipra_oscillatory_2020,avni_non-reciprocal2023} for related spin models exhibiting spontaneous magnetization oscillations.
Such oscillations lead in particular to a continuous support $[-m_0,m_0]$, with $m_0>0$, of the probability distribution of the magnetization, at odds with usual paramagnetic or ferromagnetic phases.
\modif{It is also important to note that although individual spin systems may exhibit temporal oscillations, the nonequilibrium probability distribution $P(\{s_i\},t)$ of the spin configuration $\{s_i\}$ at time $t$ converges at large time to a stationary probability distribution $P_s(\{s_i\})$ which describes a nonequilibrium ensemble of oscillating spin systems with randomly distributed phases.}

\subsection{Overlap between spin configurations}
A detailed characterization of the phase transition to an oscillating state is obtained by considering the statistics of the overlap
\be
q = \frac{1}{N} \sum_{i=1}^N s_i s_i' \qquad 
\ee
between two spin configurations $\{s_i\}$ and $\{s_i'\}$ (note that $-1\le q \le 1$).
Identical (opposite) spin configurations have an overlap $q=1$ ($q=-1$),
while $q=0$ for uncorrelated spin configurations.
To describe the overlap statistics, we introduce the overlap probability distribution
    \begin{equation}
      \label{eq:Poverlap}
        P(q)=\sum_{\{s_i\}, \{s_i'\}} P_s(\{s_i\})P_s(\{s_i'\}) \,\delta\bigg(\frac{1}{N}\sum_{i=1}^N s_i s_i'-q\bigg),
    \end{equation}
obtained by averaging over two statistically independent spin configurations $\{s_i\}$ and $\{s_i'\}$.
The overlap distribution $P(q)$ can be evaluated for $N\to \infty$, based on the spin-configuration distribution $P_s(\{s_i\})$.
As the spins are exchangeable random variables in mean-field models \modif{(i.e., statistical properties of spin configurations are invariant under an arbitrary permutation of spin labels)}, de Finetti's representation theorem \cite{hewitt_symmetric_1955,aldous_ecole_1985}
leads for large $N$ to
     \begin{equation} \label{eq:deFinetti}
        P_s(\{s_i\}) = \int_{-1}^1 \! \mathrm{d}m\, \tilde{P}(m)\, \mathcal{P}(\{s_i\}|m)
    \end{equation}
with a factorized conditional distribution $\mathcal{P}(\{s_i\}|m)$,
     \begin{equation} \label{eq:conditional:dist}
       \mathcal{P}(\{s_i\}|m) =
       \left(\frac{1-m^2}{4}\right)^{N/2} \prod_{i=1}^N \left( \frac{1+m}{1-m} \right)^{s_i/2}
     \end{equation}
and where $\tilde{P}(m)$ is the probability density of the magnetization. 
Taking a Fourier transform of Eq.~(\ref{eq:Poverlap}) and using Eqs.~(\ref{eq:deFinetti}) and (\ref{eq:conditional:dist}) (see Appendix~\ref{app:overlap:dist} for details), the following integral expression of the overlap distribution $P(q)$ is obtained, for $|q| \le 1$, as
\be \label{eq:overlap_result}
P(q) = 2\int_{|q|}^1 \frac{dm}{m}\tilde{P}(m)\tilde{P}\left(\frac{q}{m}\right) .
\ee
Hence the overlap distribution $P(q)$ is uniquely determined by the magnetization distribution $\tilde{P}(m)$.
In the presence of spontaneous oscillations of the magnetization, $\tilde{P}(m)$ can be obtained, in the infinite $N$ limit, from the shape of the limit cycle describing oscillations in the plane ($m$,$\dot{m}$), where
$\dot{m}=dm/dt$. The shape of the limit cycle is described by an equation $\dot{m}=F(m)$ for $\dot{m}>0$, where the function $F(m)$ is defined over the interval $[-m_0,m_0]$. Assuming that the reversal symmetry
$(m,\dot{m})\to (-m,-\dot{m})$ holds, the limit cycle is symmetric with respect to the origin $(m,\dot{m})=(0,0)$, and the part corresponding to $\dot{m}<0$
is described by $\dot{m}=-F(-m)$. The magnetization distribution is obtained as
\modif{
\be \label{eq:P0:m}
\tilde{P}(m) = \frac{1}{\tau} \int_{t_0}^{t_0+\tau} dt \, \delta\big( m(t)-m \big)
\ee
where $\tau$ is the oscillation period, $t_0$ is an arbitrary time, and $m(t)$ is a trajectory of the magnetization along the limit cycle, i.e., a solution of the equations $\dot{m}=F(m)$ for $\dot{m}>0$
and $\dot{m}=-F(-m)$ for $\dot{m}<0$. Integrating out the delta function, one obtains
\be \label{eq:P0:m}
\tilde{P}(m) = \frac{1}{\tau}\left( \frac{1}{F(m)} + \frac{1}{F(-m)} \right).
\ee
Hence the magnetization distribution} $\tilde{P}(m)$ is symmetric: $\tilde{P}(m)=\tilde{P}(-m)$.
The normalization of $\tilde{P}(m)$ implies that the period $\tau$ is given by $\tau=2\int_{-m_0}^{m_0} dm/F(m)$.
The shape of the limit cycle in the plane ($m$,$\dot{m}$) thus fully determines the magnetization distribution $\tilde{P}(m)$, and thus the overlap distribution $P(q)$. In other words, by monitoring the shape of the limit cycle, one may be able to reach a prescribed overlap distribution.

\subsection{Magnetization-dependent frequency}

There are many ways to monitor the shape of the limit cycle by tuning the macroscopic dynamics. In the following, we explore more specifically
the generic idea of time reparameterization, which can be simply stated as follows. As a reference model, we start from a driven mean-field spin model that exhibits spontaneous oscillations of period $\tau_0$, with a limit cycle described by $\dot{m}=F_0(m)$ for $\dot{m}>0$. Multiplying the microscopic transition rates of the spin-flip dynamics by a magnetization-dependent factor $\nu(m)$, the dynamics of the magnetization $m$ is obtained from the original dynamics through a (generally non-linear) reparametrization of time $t \to t'$, with $dt'=\nu(m) dt$, such that $m(t)=m_0(t')$, where $m_0(t)$ stands for the time-dependent magnetization in the reference model.
As a result, the shape of the limit cycle \modif{in the $(m,\dot{m})$-plane} is changed into $\dot{m} = \nu(m)F_0(m)$, \modif{with $\dot{m}>0$.
In other words, $F(m)=\nu(m)F_0(m)$,} leading according to Eq.~(\ref{eq:P0:m}) to a modified magnetization distribution
\be \label{eq:Pnu:m}
\tilde{P}(m) = \frac{1}{\tau} \left( \frac{1}{\nu(m)F_0(m)} + \frac{1}{\nu(-m)F_0(-m)} \right).
\ee
\modifnew{The choice of the magnetization-dependent frequency $\nu(m)$ is arbitrary, provided the function $\nu(m)$ remains positive (it can only vanish on isolated points).
Note that a motivation for this particular modification of the dynamics is the fact that changing the attempt frequency of the stochastic dynamics as a function of a global observable is a very generic type of change, which does not rely on any specificity of the original dynamics, and which can thus be implemented on any type of stochastic dynamics.}

\modifnew{We now try to determine the magnetization-dependent frequency $\nu(m)$ in terms of the magnetization distribution $\tilde{P}(m)$.}
The period $\tau$ is arbitrary, as it can be changed by a rescaling of $\nu(m)$, and we thus set $\tau=\tau_0$.
Under the assumption $\nu(m)=\nu(-m)$, Eq.~(\ref{eq:Pnu:m}) can then be inverted to give
\be \label{eq:nu:PmPm}
\nu(m) = \frac{\tilde{P}_0(m)}{\tilde{P}(m)},
\ee
where $\tilde{P}_0(m)$ is the magnetization distribution for the reference model.
Hence, if one is able to determine the magnetization distribution which is a solution of Eq.~(\ref{eq:overlap_result}) for a prescribed overlap distribution $P(q)$, it is then straightforward to determine the dimensionless frequency $\nu(m)$ to be included in the microscopic dynamics.
This result provides a direct link between the overlap distribution and the microscopic spin dynamics.

\subsection{Determination of $\tilde{P}(m)$ from $P(q)$}

\subsubsection{Mapping to a convolution problem}

We now discuss the generic issue of inverting the integral equation (\ref{eq:overlap_result}) to obtain $\tilde{P}(m)$ from the knowledge of $P(q)$.
\modif{We start by mapping Eq.~(\ref{eq:overlap_result}) to a more standard convolution problem, for which inversion techniques are available.}
The magnetization distribution $\tilde{P}(m)$ is assumed to have a bounded continuous support $[-m_0,m_0]$, leading for the overlap distribution $P(q)$ to a continuous support $[-q_0,q_0]$, with $q_0=m_0^2$.
Introducing the function $g(u)$ defined for $u=\ln |m_0/m|>0$ through the relation
\be
\tilde{P}(m)=m_0^{-1}g\left(\ln\left|\frac{m_0}{m}\right|\right)\theta(m_0-|m|)
\ee
and similarly the function $\phi(v)$ defined for $v=\ln |q_0/q|>0$ by 
\be
P(q)=q_0^{-1}\phi\left(\ln\left|\frac{q_0}{q}\right|\right)\theta(q_0-|q|),
\ee 
Eq.~(\ref{eq:overlap_result}) takes a convolution form,
\be \label{eq:relation:phi:g}
\phi(v)=2\int_v^1 du\,  g(u)g(v-u)\,.
\ee 
This equation is supplemented by the normalization constraints on $P(q)$ and $\tilde{P}(m)$, which in terms of the functions $g(u)$ and $\phi(v)$ become
\be \label{eq:phi:normed:condition}
2\int_0^{\infty} dv\, e^{-v}\phi(v)=1, \qquad 2\int_0^{\infty} du\, e^{-u}g(u)=1\,.
\ee

\subsubsection{Gamma-like parameterization}

The problem of finding the magnetization distribution $\tilde{P}(m)$ corresponding to a prescribed overlap distribution $P(q)$ is equivalent to starting from a prescribed function $\phi(v)$, non-zero for $v>0$ only, and inverting Eq.~(\ref{eq:relation:phi:g}) to obtain the function $g(u)$. 
One possible technique is to solve the convolution problem using a Laplace transform, and to invert numerically the Laplace transform of $g(u)$.
Alternatively, one may consider classes of functions $\phi(v)$ for which the solution $g(u)$ of Eq.~(\ref{eq:relation:phi:g}) is known analytically.
For instance, if $\phi(v)$ is proportional to an infinitely divisible distribution \cite{sato-inf-divisible2013} with support on the positive real axis and satisfies the normalization condition
(\ref{eq:phi:normed:condition}), a solution of Eq.~(\ref{eq:relation:phi:g}) can easily be found.
A simple example of an infinitely divisible distribution is the class of gamma distributions. 
To satisfy the normalization condition (\ref{eq:phi:normed:condition}), we take for $\phi(v)$ a modified gamma distribution,
\be
\phi(v) = f(v;\alpha, \beta) \equiv \frac{\beta^{\alpha}v^{\alpha-1}e^{-(\beta-1) v}}{2\Gamma(\alpha)},
\ee
with two parameters $\alpha>0$ and $\beta>0$.
The function $g(u)$ solution of Eq.~(\ref{eq:relation:phi:g}) is then given by
\be
g(u) = f\left(u; \frac{\alpha}{2}, \beta\right).
\ee
In the original variables $q$ and $m$, the overlap distribution is given by
\be \label{eq:pq:gamma}
P_{\alpha\beta}(q)=\frac{\beta^{\alpha}}{2q_0\Gamma(\alpha)}\left|\frac{q}{q_0}\right|^{\beta-1}\left(\ln\left|\frac{q_0}{q}\right|\right)^{\alpha-1}\theta(q_0-|q|),
\ee
while the magnetization distribution reads as 
\be \label{eq:pm:gamma}
\tilde{P}_{\alpha\beta}(m)=\frac{\beta^{\alpha/2}}{2m_0\Gamma(\frac{\alpha}{2})}\left|\frac{m}{m_0}\right|^{\beta-1}\!\left(\ln\left|\frac{m_0}{m}\right|\right)^{\alpha/2-1}\!\!\theta(m_0-|m|)
\ee
(we recall that $q_0=m_0^2$).
In practice, once the magnetization distribution $\tilde{P}(m)$ associated with the prescribed overlap distribution $P(q)$ has been obtained, one needs to find the values of the control parameters of the reference model (i.e., the model with $\nu(m)=1$) for which the support of $\tilde{P}_0(m)$ is the same as $\tilde{P}(m)$. 
The dimensionless frequency $\nu(m)$ is then obtained from Eq.~(\ref{eq:nu:PmPm}). 
Multiplying the transition rates of the reference model by $\nu(m)$, one obtains a model exhibiting the prescribed overlap distribution.

The simple parameterization (\ref{eq:pq:gamma}) of the overlap distribution already exhibits a rich behavior.
At a qualitative level, $\alpha$ characterizes the behavior of $P(q)$ close to the support boundaries, $P(q) \sim (q_0-|q|)^{\alpha-1}$ for $q \to \pm q_0$.
For $\alpha=1$, $P(q)$ goes to a finite, nonzero limit for $q \to \pm q_0$, similarly to the behavior observed for an elliptic limit cycle in the plane ($m$, $\dot{m}$),
see \cite{guislain_nonequil2023} and Fig.~\ref{fig:p0mp0q}.
For $\alpha>1$, $P(q)\to 0$ for $q \to \pm q_0$, while for $\alpha<1$, $P(q)$ diverges in $\pm q_0$.
A divergence of $P(q)$ at the boundaries of the support is reminiscent of the low-temperature behavior of mean-field spin-glass models, although the functional form of the divergence is quantitatively different here.
The parameter $\beta$ characterizes the behavior of $P(q)$ for $q\to 0$, $P(q) \sim |q|^{\beta-1}$ up to logarithmic corrections.
For $\beta=1$ and $\alpha>1$, a logarithmic divergence of $P(q)$ is observed for $q\to 0$, similarly to the behavior observed for an elliptic limit cycle \cite{guislain_nonequil2023}. For $\beta \ne 1$, one has for $q\to 0$ either a power-law divergence of $P(q)$ [$\beta<1$], or a zero limit [$\beta>1$].
Hence, the introduction of a reparameterization through $\nu(m)$ shows that neither the logarithmic divergence at $q=0$ nor the finite non-zero limit at the boundaries of the support of $P(q)$ are hallmarks of the presence of a limit cycle. These properties are specific to elliptic-like limit cycles
\cite{guislain_nonequil2023}.

\section{Explicit stochastic spin model}

\subsection{Definition of the model}

We now illustrate the above generic results on a specific example of a driven mean-field spin model.
We consider a generalization of the spin model studied in \cite{guislain_nonequil2023}, where spins are coupled to dynamic fields, allowing for the onset of spontaneous collective oscillations in the parameter regime when spins tend to align with fields while fields tend to antialign with spins.
The model involves $2N$ microscopic variables: $N$ spins $s_i=\pm1$ and $N$ fields $h_i=\pm 1$. 
Global observables are the magnetization $m$ and the average field $h=N^{-1} \sum_{i=1}^N h_i$.
The stochastic dynamics consists in randomly flipping a single spin $s_i$ or a single field $h_i$. The flipping rates $W_s$ and $W_h$ depend only on $m$ and $h$, 
\be \label{eq:transition:rate}
W_{s,h} =\nu(m) [1+\exp(\beta \Delta E_{s,h})]^{-1},
\ee
with $\beta=T^{-1}$ the inverse temperature and $\Delta E_{s,h}$ the variation of $E_{s,h}$ when flipping a spin $s_i$ or a field $h_i$, where
\be
E_s=-N \left(\frac{J_1}{2} m^2+\frac{J_2}{2} h^2+mh \right), \quad E_h=E_s+\mu Nhm,
\ee
and $\nu(m)$ is a magnetization-dependent microscopic frequency scale, assumed to be an even function of $m$.
We note $T_c=(J_1+J_2)/2$ the critical temperature and $\mu_c=1+(J_1-J_2)^2/4$. For $\mu>\mu_c$, the model exhibits a transition at $T_c$ from a paramagnetic phase ($T>T_c$) to an oscillating phase ($T<T_c$) \cite{guislain_nonequil2023}.

In the limit of infinite system size, one gets the following deterministic equations for the dynamic of $m(t)$ and $h(t)$, 
\be \label{eq:eq:deter}
\begin{aligned}
\frac{dm}{dt}&=\nu(m)[-m+\tanh(\beta J_1 m+\beta h)] \\ 
\frac{dh}{dt}&=\nu(m)[-h+\tanh(\beta J_2 h+\beta (1-\mu)m)].
\end{aligned}
\ee
The function $\nu(m)$ controls the local time scale of the dynamics, and it can be reabsorbed into a non-linear reparametrization of time
$t \to t'$ such that $dt'=\nu(m) dt$. Rewriting Eqs.~(\ref{eq:eq:deter}) in terms of the dimensionless time variable $t'$, the frequency $\nu(m)$ no longer explicitly appears in the dynamics. This means in particular that the shape of the limit cycle in the ($m$, $h$)-plane is independent of the functional form of $\nu(m)$. However, the local speed along the limit cycle does depend on $\nu(m)$.
\modifnew{The idea of changing the speed along the limit cycle in the $(m,h)$ representation is the key motivation for the introduction of the magnetization-dependent frequency $\nu(m)$.}
Reexpressing the dynamics in terms of the generic variables ($m$, $\dot{m}$), one finds that the shape of the limit cycle in these variables depends on $\nu(m)$.

\modifnew{To sum up, there are two different ways to consider the effect of the introduction of the magnetization-dependent frequency $\nu(m)$.
In terms of the generic variables $(m,\dot{m})$, changing $\nu(m)$ is only one possible way to change the shape of the limit cycle in the $(m,\dot{m})$-plane.
By contrast, changing $\nu(m)$ in the dynamics of the model-dependent variables $(m,h)$ does not modify the shape of the limit cycle in the $(m,h)$-plane,
but only the speed along the cycle. In this case, the effect of $\nu(m)$ can be interpreted as a time reparametrization.
This is clear from Eq.~(\ref{eq:eq:deter}), where $\nu(m)$ appears as a prefactor in the rhs of both evolution equations.}

\subsection{Overlap distribution}

\begin{figure}[t]
    \centering
    \includegraphics{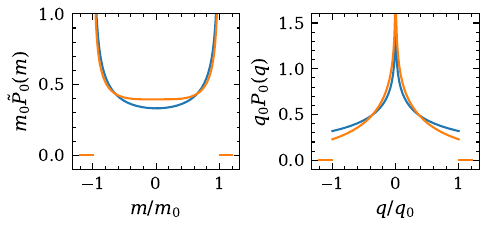}
    \caption{(a) Magnetization distribution $\tilde{P}_0(m)$, and (b) overlap distribution $P_0(q)$, for $\mu>\mu_0$ (blue/dark color) and $\mu=\mu_0$ (orange/light color), upon rescaling by the size of their support.}
    \label{fig:p0mp0q}
\end{figure}

We first consider the case $\nu(m)=1$ and denote as $\tilde{P}_0(m)$ and $P_0(q)$ the corresponding magnetization and overlap distributions.
We note $\ve=(T_c-T)/T_c$ and $\mu_0=1-(T-J_1)(T-J_2)$.
At the transition from a paramagnetic phase to an oscillating phase (small $\ve>0$), one finds for $\mu>\mu_0$ that
$q_0=m_0^2=\ve a$ [see Eq.~(\ref{eq:a}) in Appendix~\ref{app:coeffs} for the expression of $a$] and
\be
\tilde{P}_0(m) = \frac{1}{\pi m_0} \, \left[1-\left(\frac{m}{m_0}\right)^2 \right]^{-1/2} \theta(m_0-|m|).
\ee
The overlap distribution is given by $P_0(q)=q_0^{-1} \psi_0(q/q_0)$, with 
\begin{equation} \label{eq:psi:scalfn}
      \psi_0(y)= \frac{2}{\pi^2} \int_{|y|}^{1} \frac{\mathrm{d}x}{\sqrt{(1-x^2)(x^2-y^2)}}\,\theta(1-|y|).
\end{equation}
The overlap distribution has a logarithmic divergence for $q\to0$, and a non-zero limit at the boundaries of its support,
$P_0(\pm q_0) = \frac{1}{\pi q_0}$ for $|q|\to q_0$ \cite{guislain_nonequil2023}.
Close to the tricritical point, for $\mu=\mu_0$ and $\ve>0$, the magnetization still oscillates but its probability distribution
$\tilde{P}_0^{\mathrm{tc}}(m)$ takes a different form. One finds $m_0^2=b\ve$ [with $b$ given in Eq.~(\ref{eq:b}) in Appendix~\ref{app:coeffs}] and
\be
\tilde{P}_0^{\mathrm{tc}}(m) = \frac{d}{m_0} \left|1-\left(\frac{m}{m_0}\right)^4\right|^{-1/2}\theta(m_0-|m|),
\ee
with $d=\sqrt{2}\Gamma(3/4)^2/\pi^{3/2}$.
The overlap distribution $P_0^{\mathrm{tc}}(q)$ is then given by
\begin{equation} \label{eq:psi:scalfn:nonelliptic}
      P_0^{\mathrm{tc}}(q) = \frac{2d^2}{q_0} \int_{|q|/q_0}^{1} \frac{\mathrm{d}x\, x}{\sqrt{(1-x^4)[x^4-(q/q_0)^4]}}\, \theta(q_0-|q|),
\end{equation}
with $q_0=m_0^2$.
We plot in Fig.~\ref{fig:p0mp0q} the overlap distributions $P_0(q)$ and $P_0^{\mathrm{tc}}(q)$, corresponding to the cases $\mu>\mu_0$ and $\mu=\mu_0$ respectively, for small $\ve$.
Both overlap distributions have a logarithmic divergence for $q\to0$, and have a non-zero limit at the boundaries of their support, in $|q|\to m_0^2$,
\be
P_0(\pm q_0) = \frac{1}{\pi q_0}\,, \quad
P_0^{\mathrm{tc}}(\pm q_0)=\frac{\Gamma(\frac{3}{4})^4}{\pi^2 q_0}\,.
\ee

\begin{figure}[t]
    \centering
    \includegraphics{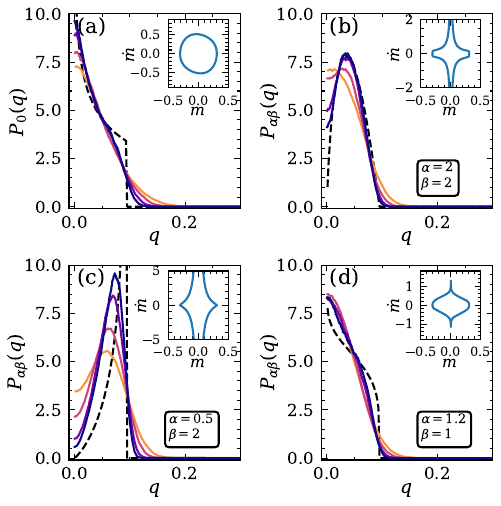}
    \caption{(a) Overlap distribution $P_0(q)$ (full lines) measured in a stochastic simulation of the spin model of size $N$ and theoretical prediction in the $N\to\infty$ limit (dashed line) from Eq.~(\ref{eq:psi:scalfn}). 
    (b)-(d) Overlap distribution $P(q)$ (full lines) measured in a stochastic simulation of the spin model of size $N$, with a frequency $\nu(m)$ determined from a prescribed distribution $P_{\alpha\beta}(q)$ (dashed lines), as given in Eq.~(\ref{eq:pq:gamma}).
    (b) $\alpha=2$, $\beta=2$; (c) $\alpha=0.5$, $\beta=2$; (d) $\alpha=1.2$, $\beta=1$. 
    Convergence to the prescribed overlap distribution is observed by increasing the system size $N$ (in both panels, $N=1000$, $2000$, $5000$ and $10000$, from lighter to darker colors).
    Inset: limit cycle in the phase space ($m, \dot{m}$) in the deterministic limit $N\to \infty$.
    Parameters: $T=0.45$, $\mu=2$, $J_1=0$, $J_2=1$.}
    \label{fig:example:pq}
\end{figure}

We now simulate the stochastic spin dynamics to numerically determine the overlap distribution $P(q)$ from the microscopic dynamics in a finite size system.
The overlap distribution $P_0(q)$ measured from stochastic simulations of the spin model of size $N$ for $\nu(m)=1$ is plotted in Fig.~\ref{fig:example:pq}(a). The convergence to the theoretical limit (dashed line) is observed by increasing the system size $N$. 
We then turn to the case when $\nu(m)$ is given by Eq.~(\ref{eq:nu:PmPm}) with $\tilde{P}(m)=\tilde{P}_{\alpha\beta}(m)$ [Eq.~(\ref{eq:pm:gamma})]
associated with a prescribed overlap distribution $P_{\alpha\beta}(q)$ as given in Eq.~(\ref{eq:pq:gamma}). We find that by increasing system size, the numerically measured overlap distributions $P(q)$ converges to the prescribed overlap distribution $P_{\alpha\beta}(q)$ [Fig.~\ref{fig:example:pq}(b-d)].
The corresponding limit cycles in the plane ($m$, $\dot{m}$) are displayed in the insets of Fig.~\ref{fig:example:pq}(b-d).
For $\alpha=1.2$ and $\beta=1$, the overlap distribution has a slow divergence in $q=0$ as $(-\ln |q|)^{0.2}$ and converges to zero at the boundary of the support. The convergence to the prescribed overlap distribution $P_{\alpha\beta}(q)$ for these parameters is slower than for other parameters shown in Fig.~\ref{fig:example:pq}.

\section{Interpretation in terms of pure states}

We finally discuss the physical interpretation of the fact that the overlap probability distribution is spread over a continuous interval,
a property reminiscent of the full replica symmetry breaking scenario in the context of disordered systems like spin-glasses \cite{mezard_spin_1987}.
In the spin models considered here, a replica symmetry breaking is not expected since no disorder is present.
\modif{An important difference with spin-glasses is also that the overlap distribution depends only on the magnetization distribution, which is not the case for spin-glasses.}
Yet, \modif{qualitative analogies} with disordered systems arise if one thinks in terms of pure states. Indeed, the standard interpretation of replica symmetry breaking
is that \modif{the equilibrium Gibbs distribution can be decomposed as a mixture of} a large number of pure states, 
for which connected correlations vanish at large distance \cite{mezard_spin_1987}.
The average value of an observable $O$ is thus given by an average over pure states, $\la O \ra = \sum_{\alpha} w_{\alpha} \la O \ra_{\alpha}$,
where $w_{\alpha}$ is the probability weight of pure state $\alpha$, and $\la O \ra_{\alpha}$ is the corresponding pure state average of the observable $O$
\cite{mezard_spin_1987}.
Although no disorder is present, the situation in spin models with spontaneous magnetization oscillations \modif{shares some qualitative analogies with spin glasses.}
\modif{As shown in Eq.~(\ref{eq:deFinetti}), the nonequilibrium steady-state distribution $P_s(\{s_i\})$ can be decomposed, thanks to the exchangeability of spin configurations, as a mixture of factorized distributions $\mathcal{P}(\{s_i\}|m)$,
defined in Eq.~(\ref{eq:conditional:dist}), that play a role similar to pure states.
These factorized states are labeled by the magnetization $m$, and have a probability weight $\tilde{P}(m)$.
They satisfy the condition that connected correlations vanish at large distance (they actually vanish for all distances as the distribution if factorized).}
The average of an arbitrary observable $O(\{s_i\})$ is then obtained as
\be
\langle O \rangle = \int_{-1}^1 dm\, \tilde{P}(m) \langle O \rangle_m
\ee
with the `pure-state' average
\be
\langle O \rangle_m = \sum_{\{s_i\}} \mathcal{P}(\{s_i\}|m) \, O(\{s_i\}).
\ee
At odds with spin-glasses, though, pure states are not organized in a tree-like, ultrametric structure \cite{mezard_spin_1987}.
Defining the distance $d_{ab}$ between two pure states $a$ and $b$ as $d_{ab}=|m_a-m_b|$
the distances between three pure states $a$, $b$ and $c$ satisfy a triangle inequality $d_{ab} \le d_{ac}+d_{cb}$, confirming the absence of
ultrametricity (note that in a spin-glass context, the Hamming distance $d_{ab}=(1-q_{ab})/2$ is often considered; here, $q_{ab}=m_a m_b$, and the Hamming distance would be nonzero for $m_a = m_b$, which would contradict the fact that there is a single pure state with a given magnetization $m$). In addition, the self-overlap $q_{aa}=m_a^2$ is not the same for all pure states, again at variance with the standard mean-field spin-glass scenario \cite{mezard_spin_1987}.

\modif{These important differences rely on the type of random-variable exchangeability at play. In the present nonequilibrium mean-field spin models, spin configurations are exchangeable, meaning that their statistical properties are invariant under an arbitrary permutation of spin indices. The exchangeability property leads to a decomposition of the nonequilibrium stationary distribution $P_s(\{s_i\})$ as a mixture of factorized distributions, according to de Finetti's representation theorem \cite{hewitt_symmetric_1955,aldous_ecole_1985}.
In mean-field spin-glass models, spins configurations are not exchangeable due to the heterogeneity introduced by disorder (e.g., quenched random coupling constants).
Yet, the spin-glass physics, and in particular the decomposition over pure states, has been shown to rely, from a mathematical standpoint, on more involved notions of exchangeability like exchangeable arrays \cite{Panchenko13} and their hierarchical generalizations \cite{Panchenko15,Austin14}, that apply to the joint set of spin and coupling constant random variables.
Hence, in this mathematical sense, nonequilibrium mean-field spin models exhibiting oscillations only borrow to spin-glasses the minimal ingredient (i.e., the generic notion of exchangeability of random variables) allowing for a nontrivial overlap distribution, while keeping a much simpler form than disordered spin models.}

\section{Conclusion}

To sum up, we have shown that the overlap distribution, a highly non-trivial observable unveiling complex statistical states, can be tailored to a broad class of prescribed distributions in driven mean-field spin models exhibiting spontaneous oscillations of the magnetization.
The overlap distribution can be directly traced back to the shape of the limit cycle in the plane ($m$, $\dot{m})$ of the magnetization and its time derivative, while the limit cycle shape can be prescribed by monitoring in a simple way the microscopic spin dynamics.
This work thus provides an example of a highly non-trivial inverse problem that can be solved explicitly.
As for future work, it would be of interest to see if such an approach may be extended beyond mean-field spin models, by considering for instance finite-dimensional spin models or interacting particle models.

\acknowledgments
L. G. acknowledges funding from the French Ministry of Higher Education and Research.

\appendix 

\section{Derivation of the overlap distribution}
\label{app:overlap:dist}

We provide here a short derivation of the expression of the overlap distribution $P(q)$ given in Eq.~(\ref{eq:overlap_result}).
We introduce the Fourier transform (or characteristic function) $\chi(\omega)$ of the overlap distribution $P(q)$,
\be
\chi(\omega)=\int_{-1}^1 dq P(q) e^{i\omega q}.
\ee
Integrating the $\delta-$function of Eq.~(\ref{eq:Poverlap}) and using the expression of $P_s(\{s_i\})$ given in Eq.~(\ref{eq:deFinetti}), $\chi(\omega)$ becomes
\begin{equation}
\label{eq:chi}
    \begin{split}
\chi(\omega)= \iint dm_a dm_b \tilde{P}(m_a)\tilde{P}(m_b)e^{\frac{N}{2}\ln\left(\frac{(1-m_a^2)}{4}\frac{(1-m_b^2)}{4}\right)}\\\times 
\sum_{\{s_i^a\}, \{s_i^b\}}\prod_{i=1}^N  \left(\frac{1+m_a}{1-m_a}\right)^{\frac{s_i^a}{2}}\! \left(\frac{1+m_b}{1-m_b}\right)^{\frac{s_i^b}{2}}\!e^{\frac{i\omega}{N} s_i^as_i^b}\, .
    \end{split}
\end{equation}
Exchanging sum and product, the term on the second line of Eq.~\eqref{eq:chi} simplifies to
\be \left(\gamma_1 e^{\frac{i\omega}{N}}+\gamma_2e^{-\frac{i\omega}{N}}\right)^N\ee
with 
\be \gamma_1=\sqrt{\frac{1+m_a}{1-m_a}\frac{1+m_b}{1-m_b}}+ \sqrt{\frac{1-m_a}{1+m_a}\frac{1-m_b}{1+m_b}},\ee and 
\be \gamma_2=\sqrt{\frac{1+m_a}{1-m_a}\frac{1-m_b}{1+m_b}}+ \sqrt{\frac{1-m_a}{1+m_a}\frac{1+m_b}{1-m_b}}\,.\ee
In the large $N$ limit, it becomes
\be \left(\gamma_1e^{\frac{i\omega}{N}}+\gamma_2e^{-\frac{i\omega}{N}}\right)^N=e^{\frac{N}{2}\ln(\gamma_1+\gamma_2)^2}e^{i\omega \frac{\gamma_1-\gamma_2}{\gamma_1+\gamma_2}}\,.\ee
One can check the following identities,
\be
\frac{1}{\left(\gamma_1+\gamma_2\right)^2} = \frac{(1-m_a^2)}{4}\frac{(1-m_b^2)}{4}
\ee
and
\be
\frac{\gamma_1-\gamma_2}{\gamma_1+\gamma_2} = m_a m_b\,.
\ee
The Fourier transform of the overlap distribution $P(q)$ thus becomes 
\be
\chi(\omega)=\iint dm_a dm_b \tilde{P}(m_a)\tilde{P}(m_b) e^{i\omega m_am_b}\,.
\ee
Taking the inverse Fourier transform, one obtains 
\be   
P(q)=\iint dm_a dm_b \tilde{P}(m_a)\tilde{P}(m_b)\delta(m_am_b-q),\ee
which becomes Eq.~(\ref{eq:overlap_result}), after integration over $m_b$.

\section{Expression of the model coefficients}
\label{app:coeffs}
The explicit spin model defined in the main text involves the parameters $\ve$, $\mu_0$, $\alpha_1$ and $\alpha_2$,
whose values are given by:
\be\begin{aligned} 
&\alpha_{1}= (2 J_2 + J_2^3 + 3 J_1)/T -  2(1 + J_2^2) (-1 + J_1 J_2 +\mu)/T^2\\ &\qquad -2 + J_2 (-1 + J_1 J_2 +\mu)^2/T^3,\\
&\alpha_{3}=-2/3 +(2J_2 + J_2^3 + 3J_1)/3T.
\end{aligned}\ee
For $\mu>\mu_0$, one gets $m_0^2=a\ve$ when $\ve$ is small, with
\be \label{eq:a} a=\frac{8T_c T}{T^2\alpha_1+3(\mu-\mu_0)\alpha_3},\ee
while for $\mu=\mu_0$, one gets $m_0=b\ve$ with
\be \label{eq:b} b=\frac{5\pi^2}{3\Gamma(3/4)^4}\frac{T_c}{T\alpha_1}.\ee

\end{document}